\documentclass[epj]{webofc}\usepackage[varg]{txfonts}\usepackage{bm}
\woctitle{CONF12}
\usepackage{booktabs}\usepackage{color}\usepackage{diagbox}
\usepackage{dcolumn}\newcolumntype{d}[1]{D{.}{.}{#1}}

\begin{document}\title{Strong Couplings of Three Mesons with
Charm(ing) Involvement}\author{Wolfgang Lucha\inst{1}\fnsep\thanks
{\email{Wolfgang.Lucha@oeaw.ac.at}}\and Dmitri Melikhov
\inst{1,2,3}\fnsep\thanks{\email{dmitri_melikhov@gmx.de}}\and
Hagop Sazdjian\inst{4}\fnsep
\thanks{\email{sazdjian@ipno.in2p3.fr}}\and Silvano Simula\inst{5}
\fnsep\thanks{\email{simula@roma3.infn.it}}}\institute{Institute
for High Energy Physics, Austrian Academy of Sciences,
Nikolsdorfergasse 18, A-1050 Vienna, Austria\and D.~V.~Skobeltsyn
Institute of Nuclear Physics, M.~V.~Lomonosov Moscow State
University, 119991, Moscow, Russia\and Faculty of Physics,
University of Vienna, Boltzmanngasse 5, A-1090 Vienna, Austria\and
IPN, CNRS/IN2P3, Universit\'e Paris-Sud 11, F-91406 Orsay, France
\and INFN, Sezione di Roma Tre, Via della Vasca Navale 84, I-00146
Roma, Italy}

\abstract{We determine the strong couplings of three mesons that
involve, at least, one $\eta_c$ or $J/\psi$ meson, within the
framework of a constituent-quark model by means of relativistic
dispersion formulations. For strong couplings of $J/\psi$ mesons
to two charmed mesons, our approach leads to predictions roughly
twice as large as those arising from QCD sum rules.}\maketitle

\section{Three-meson strong coupling from meson--meson transition
amplitudes}We determine the strong couplings of three mesons at
least one of which is one of the charmonia $\eta_c$ and $J/\psi,$
generically called $g_{PP'V}$ and $g_{PV'V}$ for pseudoscalar
mesons $P$ of mass $M_P$ and vector~mesons $V$ of mass $M_V$ and
polarization vector $\varepsilon_\mu$ and defined, for momentum
transfer $q\equiv p_1-p_2,$ by the~amplitudes\begin{align*}\langle
P'(p_2)\,V(q)|P(p_1)\rangle&=-\frac{g_{PP'V}}{2}\,(p_1+p_2)^\mu\,
\varepsilon^*_\mu(q)\ ,\\\langle V'(p_2)\,V(q)|P(p_1)\rangle&=
-g_{PV'V}\,\epsilon_{\mu\nu\rho\sigma}\,\varepsilon^{*\mu}(q)\,
\varepsilon^{*\nu}(p_2)\,p_1^\rho\,p_2^\sigma\ ,\end{align*}from
the residues of \emph{poles\/} situated at the masses $M_{P_R}$
and $M_{V_R}$ of (appropriate) pseudoscalar and vector resonances
$P_R$ and $V_R$ and contributing to \emph{transition form
factors\/} $F_+^{P\succ P'}(q^2),$ $V^{P\succ V}(q^2)$ and
$A_0^{P\succ V}(q^2),$ in terms of vector quark currents
$j_\mu\equiv\bar q_1\,\gamma_\mu\,q_2$ and axial-vector quark
currents $j_\mu^5\equiv\bar q_1\,\gamma_\mu\,\gamma_5\,q_2$
defined~by\begin{align*}\langle P'(p_2)|j_\mu|P(p_1)\rangle&=
F_+^{P\succ P'}(q^2)\,(p_1+p_2)_\mu+\cdots\ ,&\left.F_+^{P\succ
P'}(q^2)\right|_{\rm pole}&=\frac{g_{PP'V_R}\,f_{V_R}}{2\,M_{V_R}
\left(1-q^2/M_{V_R}^2\right)}\ ,\\\langle
V(p_2)|j_\mu|P(p_1)\rangle&=\frac{2\,V^{P\succ V}(q^2)}{M_P+M_V}\,
\epsilon_{\mu\nu\rho\sigma}\,\varepsilon^{*\nu}(p_2)\,p_1^\rho\,
p_2^\sigma\ ,&\left.V^{P\succ V}(q^2)\right|_{\rm pole}&=
\frac{(M_V+M_P)\,g_{PVV_R}\,f_{V_R}}{2\,M_{V_R}\left(1-q^2/M^2_{V_R}
\right)}\ ,\\\langle V(p_2)|j_\mu^5|P(p_1)\rangle&={\rm i}\,
q_\mu\,(\varepsilon^*(p_2)\,p_1)\,\frac{2\,M_V}{q^2}\,A_0^{P\succ
V}(q^2)+\cdots\ ,&\left.A_0^{P\succ V}(q^2)\right|_{\rm pole}&=
\frac{g_{PP_RV}\,f_{P_R}}{2\,M_{V}\left(1-q^2/M^2_{P_R}\right)}\
,\label{T}\end{align*}where the $P$ or $V$ decay constants
$f_{P,V}$ parametrize the matrix elements of the interpolating
currents~$j_\mu^{(5)}$$$\langle0|j_\mu^5|P(q)\rangle={\rm i}\,f_P
\,q_\mu\ ,\qquad\langle0|j_\mu|V(q)\rangle=f_V\,M_V\,
\varepsilon_\mu(q)\ .$$Such strong-coupling results may prove to
be useful for studies of long-distance QCD effects in hadron
decays involving charmed mesons or charmonia in the final state of
a kind similar to the one in Ref.~\cite{PS}.

\section{Quark-model-underpinned dispersion analysis of transition
form factors}We describe the relevant properties of the involved
strongly coupling mesons by means of a relativistic
constituent-quark model \cite{CQM,CQMa,CQMb}. Of course, this
requires us to match the QCD currents $j_\mu^{(5)}$ to associated
constituent-quark currents, which is, for heavy quarks, easily
effected by introducing form factors $g_{V,A},$ $$j_\mu=g_V\,\bar
Q_1\,\gamma_\mu\,Q_2+\mbox{other Lorentz structures}\ ,\qquad
j_\mu^5=g_A\,\bar Q_1\,\gamma_\mu\,\gamma_5\,Q_2+\mbox{other
Lorentz structures}\ ,$$for which we choose $g_V=g_A=1$ \cite{MS}
but, for light quarks, rendered rather involved \cite{GVA,GVAa},
for instance, if embedding partial axial-current conservation. For
the radial meson wave functions, Gaussian shapes$$w_{P,V}(k^2)
\propto\exp\left(-\frac{k^2}{2\,\beta_{P,V}^2}\right)\,,\qquad\int{\rm
d}k\, k^2\,w^2_{P,V}(k^2)=1\ ,$$with slopes $\beta_{P,V}$ given,
together with all relevant mesonic features, in Table~\ref{P}
\cite{MP,MPa,MPb,MPc,MPd,MPe}, turn out to~suffice for our
purposes. Table~\ref{Q} lists the numerical values adopted for the
masses of the constituent quarks~$Q.$

\begin{table}[b]\centering\caption{Relevant parameters of the
mesons: numerical values of mass $M,$ leptonic decay constant $f$
and slope $\beta.$}\label{P}\begin{tabular}{ld{1.3}ld{1.3}}
\toprule Meson&\multicolumn{1}{c}{$M$~(GeV)}&\multicolumn{1}{c}
{$f$~(MeV)}&\multicolumn{1}{c}{$\beta$~(GeV)}\\\midrule
$D$&1.87&$206\pm8$&0.475\\$D^*$&2.010&$260\pm10$&0.48\\[.5ex]
$D_s$&1.97&$248\pm2.5$&0.545\\$D_s^*$&2.11&$311\pm9$&0.54\\[.5ex]
$\eta_c$&2.980&$394.7\pm2.4$&0.77\\$J/\psi$&3.097&$405\pm7$&0.68\\
\bottomrule\end{tabular}\end{table}

\begin{table}[b]\centering\caption{Constituent mass of
each quark flavour $Q=u,d,s,c$ \cite{MS} involved in charm(ing)
three-meson couplings.}\label{Q}\begin{tabular}{ld{1.2}}\toprule
Quark flavour&\multicolumn{1}{c}{Quark mass $m$\;(GeV)}\\\midrule
$u$&0.23\\$d$&0.23\\$s$&0.35\\$c$&1.45\\\bottomrule\end{tabular}
\end{table}

\begin{figure}[t]\centering
\includegraphics[scale=.2435,clip]{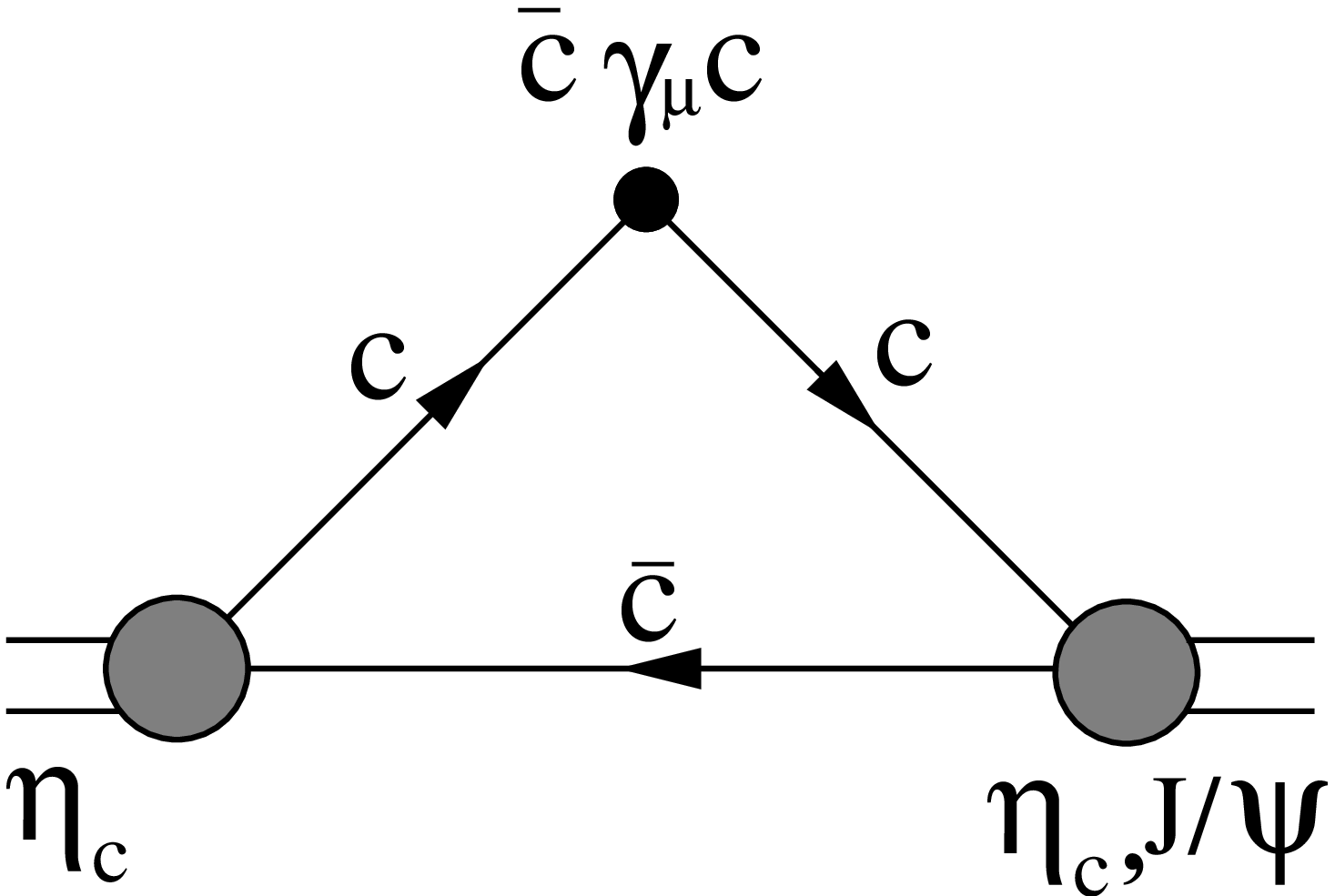}\hspace{10ex}
\includegraphics[scale=.2435,clip]{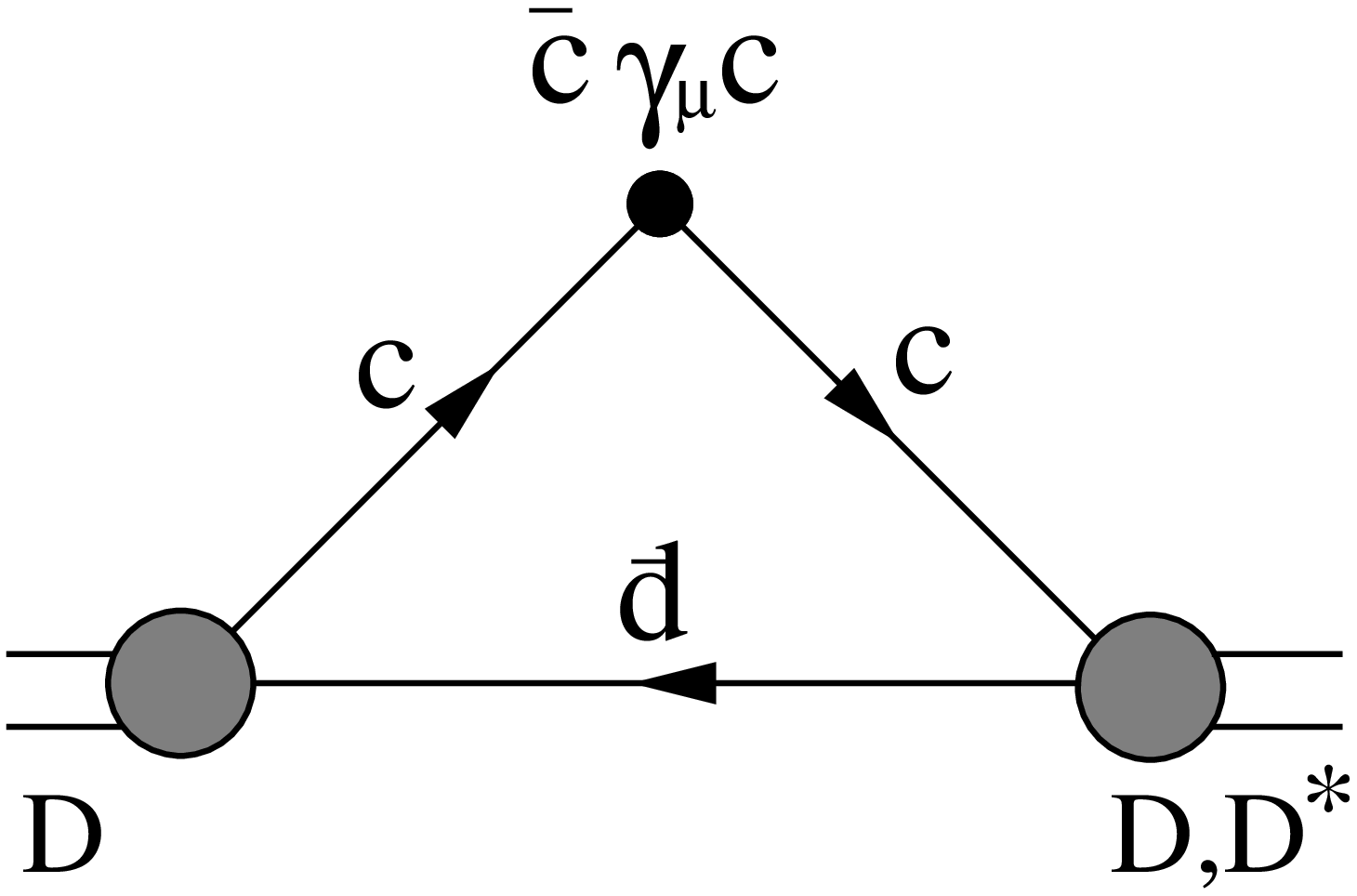}\hspace{10ex}
\includegraphics[scale=.2435,clip]{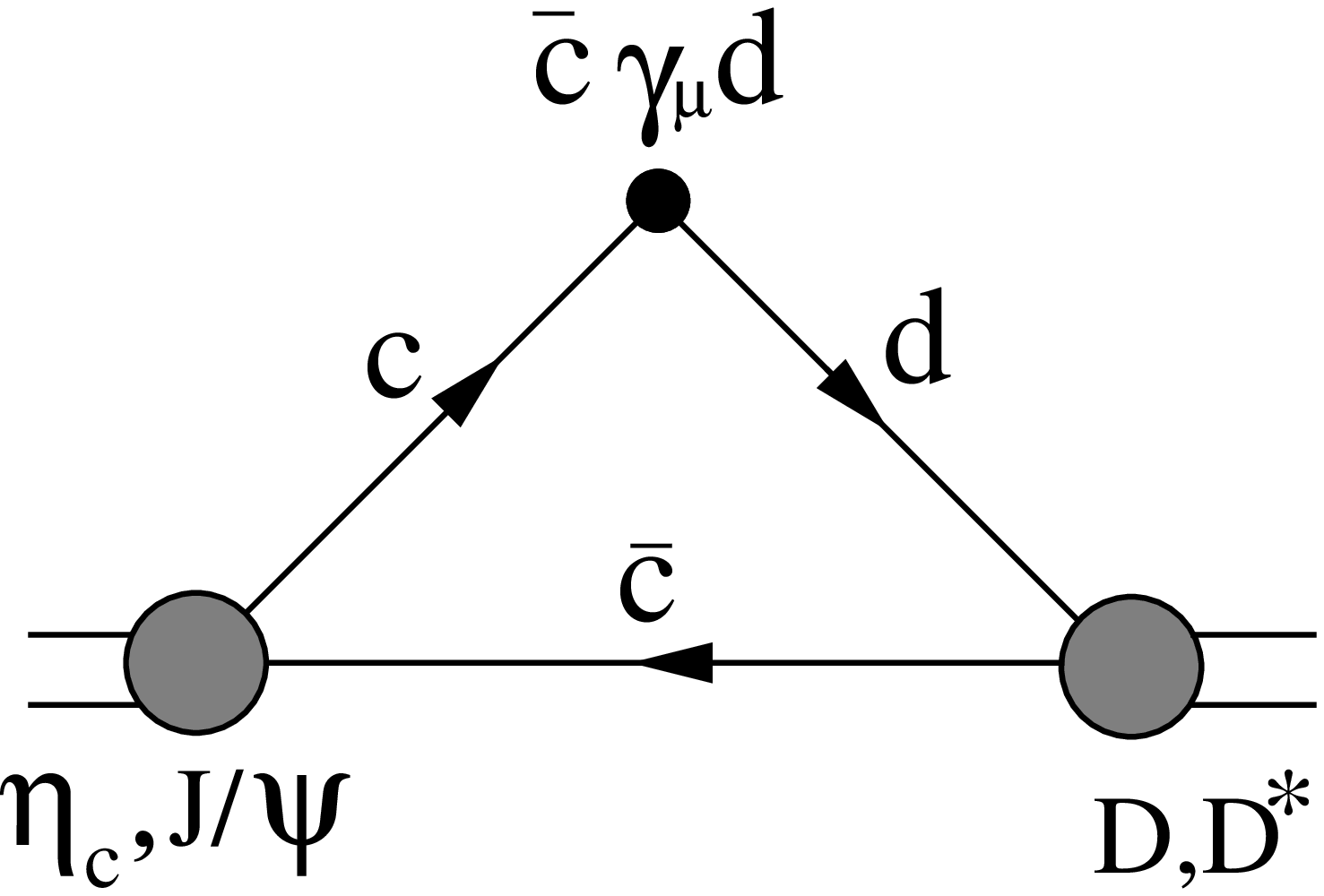}\caption{Feynman
graphs of transitions among the mesons $\eta_c,J/\psi,D,D^*,$
induced by the currents $\bar c\,\gamma_\mu\,c$ or~$\bar
c\,\gamma_\mu\,d.$}\label{FD}\end{figure}

\noindent Within the framework of a relativistic dispersion
formalism (reviewed, e.g., in Ref.~\cite{RDA}), we represent each
transition form factor ${\cal F}(q^2)=F_+^{P\succ
P'}(q^2),V^{P\succ V}(q^2),A_0^{P\succ V}(q^2)$ by a double
dispersion integral~of~a double spectral density $\Delta_{\cal
F}(s_1,s_2,q^2)$ the one-loop contributions to which derive from
Feynman graphs like the ones in Fig.~\ref{FD} and each decay
constant $f_{P,V}$ by a dispersion integral of a spectral
density~$\rho_{P,V}(s),$ $${\cal F}(q^2)=\int{\rm d}s_1\,{\rm
d}s_2\,\phi_1(s_1)\,\phi_2(s_2)\,\Delta_{\cal F}(s_1,s_2,q^2)\
,\qquad f_{P,V}=\int\limits_{(m_1+m)^2}^\infty{\rm d}s\,
\phi_{P,V}(s)\,\rho_{P,V}(s)\ ,$$and of the wave functions of all
mesons entering the corresponding one- or two-meson matrix
elements$$\phi_{P,V}(s)=\frac{\pi}{s^{3/4}}
\sqrt{\frac{s^2-(m_1^2-m^2)^2}{2\,[s-(m_1-m)^2]}}\,w_{P,V}\!
\left(\frac{(s-m_1^2-m^2)^2-4\,m_1^2\,m^2}{4\,s}\right)\,.$$

\section{Three-meson strong coupling: determination from transition
amplitudes}
\begin{figure}[b]\centering
\includegraphics[scale=.4006,clip]{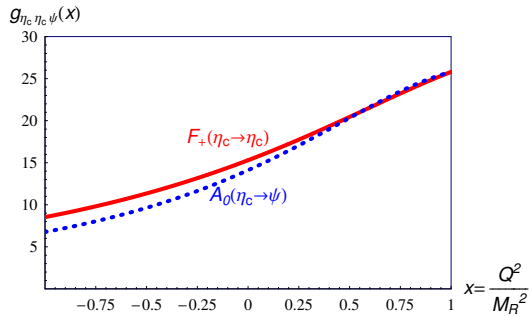}
\caption{Behaviour of the off-shell $\eta_c$-$\eta_c$-$J/\psi$
strong coupling $g_{\eta_c\eta_c\psi}$ with increasing
resonance-mass-normalized momentum transfer $x\equiv q^2/M_R^2$
for the transition of $\eta_c$ to the $\eta_c$ (solid
\textcolor{red}{red} line) or the $J/\psi$ meson (dotted
\textcolor{blue}{blue}~line).}\label{c}\end{figure}

\begin{figure}[t]\centering\begin{tabular}{ccc}
\includegraphics[scale=.466,clip]{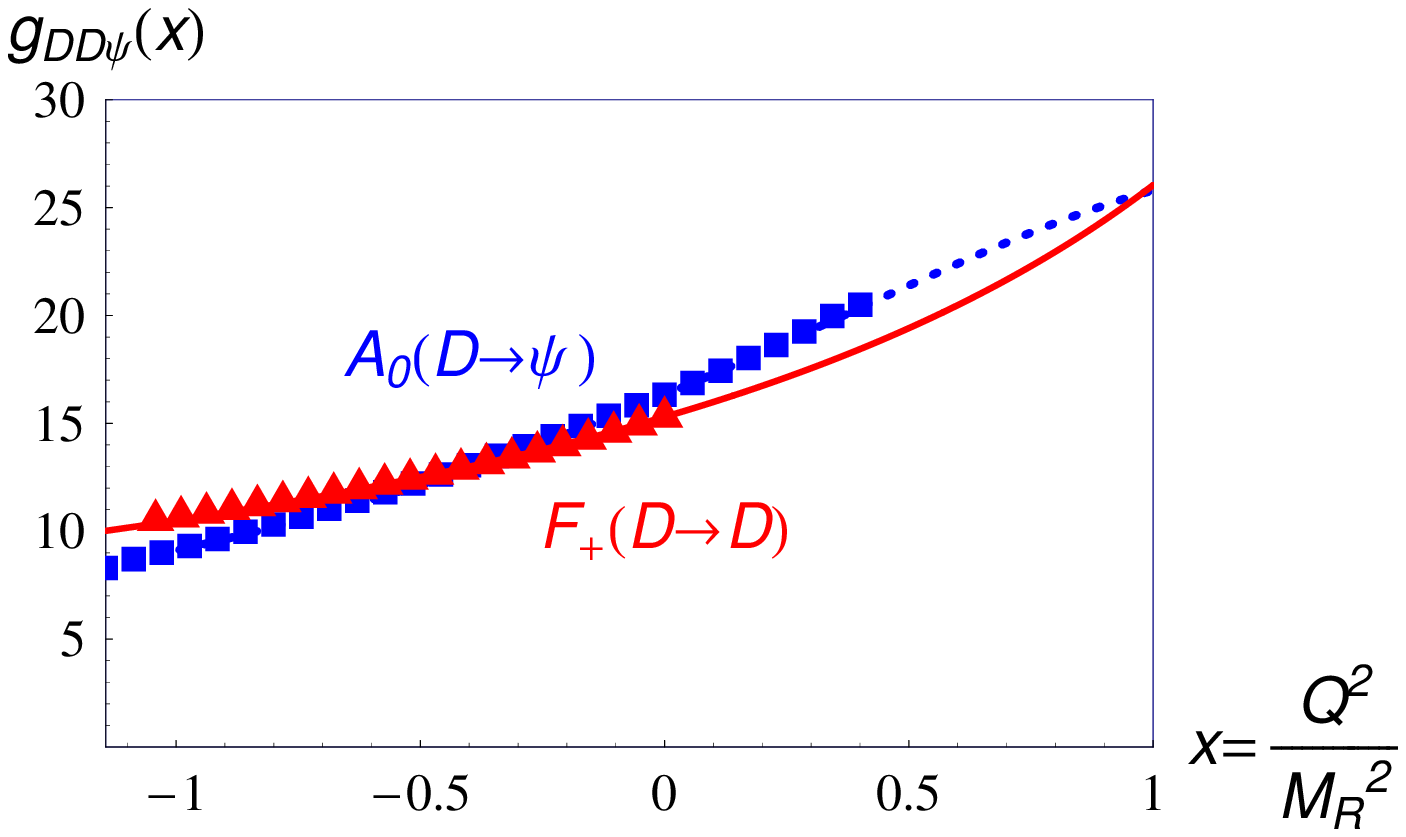}&
\includegraphics[scale=.409,clip]{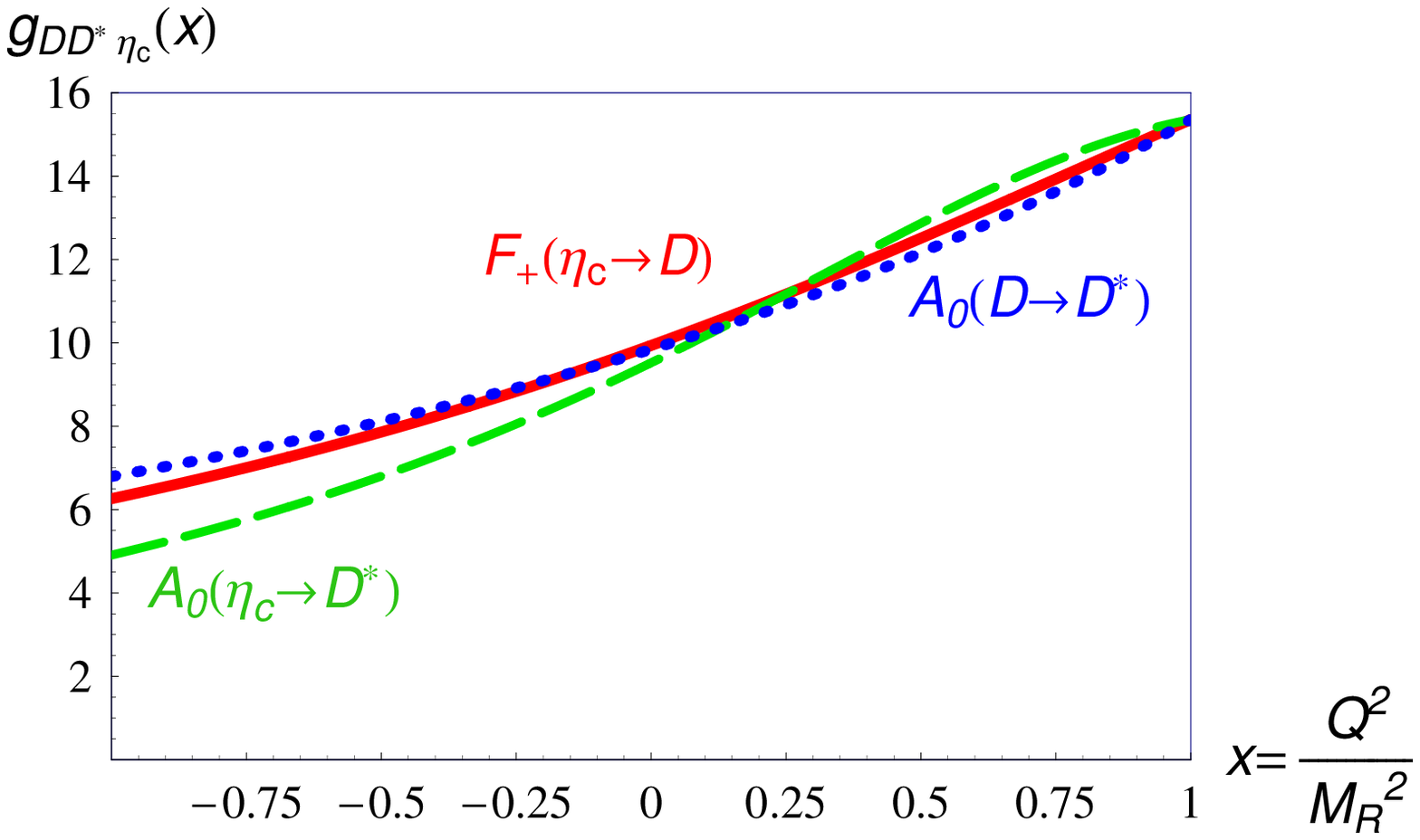}\\(a)&(b)
\end{tabular}\caption{Behaviour of the ``off-shell'' $D$-$D$-$J/\psi$
and $D$-$D^*$-$\eta_c$ strong couplings $g_{DD\psi}$ and
$g_{DD^*\eta_c},$ respectively, with increasing
``resonance-mass-normalized'' momentum transfer $x\equiv
q^2/M_R^2$: (a) $g_{D\hat D\psi}(x)=2\,M_\psi\,(1-x)\,
A^{D\succ\psi}_0(q^2)/f_D$ (\textcolor{blue}{blue} dotted line and
squares \textcolor{blue}{$\blacksquare$}) and
$g_{DD\hat\psi}(x)=2\,M_\psi\,(1-x)\,F_+^{D\succ D}(q^2)/f_\psi$
(\textcolor{red}{red} solid line and triangles
\textcolor{red}{$\blacktriangle$}), relying on interpolation
(\textcolor{blue}{blue} or \textcolor{red}{red} lines) or not
(squares \textcolor{blue}{$\blacksquare$}, triangles
\textcolor{red}{$\blacktriangle$}); (b) $g_{D\hat{D^*}\eta_c}(x)$
(solid \textcolor{red}{red} line), $g_{DD^*\hat{\eta_c}}(x)$
(dotted~\textcolor{blue}{blue} line) and $g_{\hat DD^*\eta_c}(x)$
(dashed \textcolor{green}{green} line). For each transition, the
relevant resonance, $R,$ is identified by a circumflex.}
\label{d}\end{figure}

We fix the slopes $\beta_{P,V}$ such that the decay
constants $f_{P,V}$ are reproduced by their spectral
representation. Equipped with these $\beta_{P,V}$ values, we
deduce all strong couplings from the spectral representation of
the relevant form factors ${\cal F}(q^2)$ derived sufficiently off
the resonances at $M_R$ ($R=P_R,V_R$), by interpolating pointwise
given momentum dependences of ${\cal F}(q^2)$ by three-parameter
($\sigma_{1,2},{\cal F}(0)$) ans\"atze of the form$${\cal F}(q^2)=
\frac{{\cal F}(0)}{1-\sigma_1\,q^2/M_R^2+\sigma_2\,q^4/M_R^4}\,
\frac{1}{1-q^2/M_R^2}\ ,\qquad\mbox{Res}\,{\cal F}(M_R^2)=
\frac{{\cal F}(0)} {1-\sigma_1+\sigma_2}\ ,\qquad R=P_R,V_R\
,$$and extrapolating ${\cal F}(q^2)$ to the poles at $q^2=M_R^2,$
where the strong couplings emerge from the residues. Using
$\sigma_{1,2},$ ${\cal F}(0),$ and $M_R$ as fit parameters, all
arising masses $M_R$ come close to the known resonances. Quite
generally, a given strong coupling may show up in and therefore
can be extracted from more~than one meson--meson transition form
factor, for example, $g_{\eta_c\eta_c\psi}$ from
$F_+^{\eta_c\succ\eta_c}$ or $A^{\eta_c\succ\psi}_0$ (see
Fig.~\ref{c}) \cite{LMSS,LMSSa,LMSSb}, $g_{DD\psi}$ from
$F_+^{D\succ D}$ or $A^{D\succ\psi}_0$ (see Fig.~\ref{d}(a))
\cite{LMSS,LMSSa,LMSSb} and $g_{DD^*\eta_c}$ from
$F_+^{\eta_c\succ D},$ $A^{\eta_c\succ D^*}_0$ or $A^{D\succ
D^*}_0$ (see~Fig.~\ref{d}(b)) \cite{LMSS,LMSSa,LMSSb}; for further
examples of such multiple involvements, consult Tables III, V, and
VI of Ref.~\cite{LMSS}.

\section{Strong coupling predictions from relativistic
constituent-quark approach}We collect our emerging strong-coupling
findings --- extracted, in the case of multipresence of one and
the same three-meson coupling in more than one meson--meson
transition amplitude, by a combined fit --- in Table~\ref{R}:
Strange quark content instead of a down quark implies a reduction
of the involved~strong couplings, by roughly 10\%. Confronting, in
Table~\ref{C}, our $D_{(s)}$-$D_{(s)}^{(*)}$-$J/\psi$ predictions
with QCD sum-rule outcomes \cite{Mat05,Bra14,Bra15}, the QCD
sum-rule estimates prove to be lower than ours
\cite{LMSS,LMSSa,LMSSb} by a factor~of~two.

\begin{table}[h]\centering\caption{Charm(ing) three-meson strong
couplings: quark-model-based dispersion-approach outcomes
\cite{LMSS,LMSSa,LMSSb}.}\label{R}\begin{tabular}{ld{1.8}}
\toprule$PP'V$ Coupling& \multicolumn{1}{c}{Strong coupling
$g_{PP'V}$}\\\midrule$\eta_c$-$\eta_c$-$J/\psi$&25.8\pm1.7\\[.5ex]
$D$-$D$-$J/\psi$&26.04\pm1.43\\$D$-$D^*$-$\eta_c$&15.51\pm0.45\\[.5ex]
$D_s$-$D_s$-$J/\psi$&23.83\pm0.78\\$D_s$-$D_s^*$-$\eta_c$&14.15\pm0.52
\\\bottomrule\end{tabular}
$\qquad\!\!$\begin{tabular}{ld{1.8}}\toprule$PV'V$ coupling&
\multicolumn{1}{c}{Strong coupling $g_{PV'V}$\;(GeV$^{-1}$)}\\
\midrule$\eta_c$-$J/\psi$-$J/\psi$&10.6\pm1.5\\[.5ex]
$D$-$D^*$-$J/\psi$&10.7\pm0.4\\$D^*$-$D^*$-$\eta_c$&9.76\pm0.32\\[.5ex]
$D_s$-$D_s^*$-$J/\psi$&9.6\pm0.8\\$D_s^*$-$D_s^*$-$\eta_c$&8.27\pm0.37
\\\bottomrule\end{tabular}\end{table}

\begin{table}[h]\centering\caption{Strong couplings of the
$J/\psi$ meson to two charmed mesons: relativistic quark model
vs.~QCD sum rule.}\label{C}\begin{tabular}{ld{1.13}d{1.15}r}
\toprule\diagbox{Coupling}{Approach}&\multicolumn{1}{c}{Quark
model \cite{LMSS,LMSSa,LMSSb}}&\multicolumn{1}{c}{QCD sum
rules}&\multicolumn{1}{c}{References}\\\midrule$D$-$D$-$J/\psi$&
26.04\pm1.43&11.6\pm1.8&\cite{Mat05}\\$D$-$D^*$-$J/\psi$&
(10.7\pm0.4)\;\mbox{GeV$^{-1}$}&(4.0\pm0.6)\;\mbox{GeV$^{-1}$}&
\cite{Mat05}\\[.5ex]$D_s$-$D_s$-$J/\psi$&23.83\pm0.78&
11.96^{+1.34}_{-1.16}&\cite{Bra14}\\[.5ex]$D_s$-$D_s^*$-$J/\psi$&
(9.6\pm0.8)\;\mbox{GeV$^{-1}$}&(4.30^{+1.53}_{-1.22})\;\mbox{GeV$^{-1}$}&
\cite{Bra15}\\\bottomrule\end{tabular}\end{table}

\acknowledgement{D.~M.\ would like to express gratitude for
support by the Austrian Science Fund (FWF) under project
P29028-N27.}

\end{document}